\documentclass[manuscript]{aastex}
\usepackage{lineno}
\usepackage{tikz}
\usepackage{amsmath}

\usepackage{graphicx}

\usepackage{pdflscape}

\slugcomment{\textbf{Astronomical Journal, in press} }

\shorttitle{Development of K2}
\shortauthors{Jewitt}

\begin{document}

\title{COMETARY ACTIVITY BEGINS AT KUIPER BELT DISTANCES: EVIDENCE FROM C/2017 K2}


\author{David Jewitt$^{1,2}$\\
Yoonyoung Kim$^3$, Max Mutchler$^4$, Jessica Agarwal$^3$, Jing Li$^{1}$,  \\
   Harold Weaver$^5$
} 
\affil{$^1$Department of Earth, Planetary and Space Sciences,
UCLA, 
595 Charles Young Drive East, 
Los Angeles, CA 90095-1567\\
$^2$Department of Physics and Astronomy,
University of California at Los Angeles, \\
430 Portola Plaza, Box 951547,
Los Angeles, CA 90095-1547\\
$^3$ Institut for Geophysik und Extraterrestrische Physik, Technische Universitat Braunschweig, Mendelssohnstr. 3, 38106
Braunschweig, Germany\\
$^4$ Space Telescope Science Institute, 3700 San Martin Drive, Baltimore, MD 21218 \\
$^5$ The Johns Hopkins University Applied Physics Laboratory, 11100 Johns Hopkins Road, Laurel, Maryland 20723  \\
}

\email{jewitt@ucla.edu}

\begin{abstract}
We  study the development of activity in the incoming long-period comet C/2017 K2 over the heliocentric distance range  $9 \lesssim r_H \lesssim 16$ AU.  The comet continues to be characterized by a coma of sub-millimeter and larger particles ejected at low velocity.  In a fixed co-moving volume around the nucleus we find that the scattering cross-section of the coma, $C$, is related to the heliocentric distance by a power law, $C \propto r_H^{-s}$, with heliocentric index $s = 1.14\pm0.05$. This dependence is significantly weaker than the $r_H^{-2}$ variation of the  insolation as a result of two effects.  These are, first, the heliocentric dependence of the dust velocity  and, second, a lag effect due to very slow-moving  particles ejected long before the observations were taken.     A Monte Carlo  model of the photometry shows that dust production beginning at   $r_H \sim$ 35 AU is needed to match the measured heliocentric index, with only a slight dependence on the particle size distribution.  Mass loss rates in dust at 10 AU are of order 10$^3$ kg s$^{-1}$, while loss rates in gas may be much smaller, depending on the unknown dust to gas ratio.  Consequently, the ratio of the non-gravitational acceleration to the local solar gravity, $\alpha'$, may, depending on the nucleus size, attain values $\sim$10$^{-7} \lesssim \alpha' \lesssim$ 10$^{-5}$, comparable to  values found in short-period comets at much smaller distances.  Non-gravitational acceleration in C/2017 K2 and similarly distant comets, while presently unmeasured,  may limit the accuracy with which we can infer the properties of the Oort cloud from the orbits of long-period comets.  

\end{abstract}


\section{INTRODUCTION}
\label{intro}
The dominant cometary volatile, water,  can sublimate appreciably  at blackbody temperatures prevailing out to about 5 AU, corresponding to the orbit of Jupiter  (Whipple 1950).  Comets that are active at much larger distances must be powered by the sublimation of a more  volatile material (for example, carbon monoxide or carbon dioxide) or by another physical process (for example, crystallization of amorphous ice, or non-thermal processes).   The prime  example of an inbound, distantly active comet is the long period  C/2017 K2 (Pan STARRS) (hereafter ``K2''), which was discovered  at heliocentric distance $r_H$ = 15.9 AU and  found to be active in pre-discovery data out to $r_H$ = 23.7 AU (Jewitt et al.~2017b, Meech et al.~2017, Hui et al.~2018).    Comet K2  is especially important because  its long orbital period, $\sim$3 Myr,  effectively removes the possibility that activity could be due to conducted heat trapped since the previous perihelion.  This is not true of  comets in which activity has been observed after perihelion, even at large heliocentric distances (e.g.~Prialnik and Bar-Nun 1992, Prialnik 1997, Kulyk et al.~2018).

The current heliocentric osculating orbit of K2 is marginally hyperbolic (semimajor axis $a$ = -5034 AU, perihelion distance $q$ = 1.810 AU, eccentricity $e$ = 1.00036 and inclination $i$ = 87.5\degr) with perihelion expected on UT 2022 December 20.   However, the slight excess of the eccentricity above unity  is not an indicator of an interstellar origin; numerical integrations  reveal a pre-entry, barycentric semi-major axis $a \sim$ 20,000 AU and eccentricity $e$ = 0.9998 and, as noted above, a period $\sim$3 Myr (Krolikowska and Dybczynski 2018).   The previous perihelion was nominally in the 2 AU to 4 AU range and, while it is quite uncertain, has a 97\% chance of being smaller than $q$ = 10 AU, according to these authors.  Comet K2  is thus  not dynamically new in the Oort sense, but is a return visitor  to the planetary region.  Nevertheless,  K2 is entering the planetary region of the solar system from Oort cloud distances (aphelion $\sim$ 40,000 AU)  and temperatures ($\lesssim$10 K), retaining negligible heat from the previous perihelion millions of years ago.

For these reasons, K2 is a model object in which to follow the development of activity upon approach to the Sun.  Indeed, the range of pre-perihelion distances over which K2 will be observable, from the pre-discovery observation at 24 AU to perihelion on UT 2022 December 20 at 1.8 AU, is unprecedented.  Here, we present observations obtained as part of a continuing program to monitor K2 using the Hubble Space Telescope (HST).  The  data cover the  heliocentric distance range from 15.9 AU to 8.9 AU, placing comet K2 in an essentially un-observed cometary realm.

\section{OBSERVATIONS}
We used data from the UVIS channel of  the WFC3 imaging camera on the 2.4 m Hubble Space Telescope (HST), taken  under observational programs GO 15409, 15423 and 15973.  Since the ephemeris of K2 is well-known and the size of the coma not large,  we read out only a 2058$\times$2176 pixel  subarray, giving an 81\arcsec$\times$81\arcsec~field of view at image scale 0.04\arcsec~pixel$^{-1}$.  The point-spread function from WFC3 was measured at 0.085\arcsec~FWHM (full width at half maximum), corresponding to $\sim$600 km at 10 AU.  The wide spectral response of the F350LP filter was chosen to provide maximum sensitivity to low surface brightness in the coma.  The peak system throughput is 29\% and the filter takes in most of the optical spectrum at wavelengths $\lambda >$ 3500\AA.  The effective wavelength on a sun-like spectrum source is 5846\AA~and the effective full-width at half maximum (FWHM) is  4758\AA.   

In each  HST orbit we typically obtained six integrations of 260 s each (1560 s per orbit) from which a cosmetically clean image was obtained by  computing the clipped median of the six images from each orbit after first shifting them to a common center.  The median images were also rotated to the standard orientation (north up and east left) prior to measurement.  In addition to obtaining coverage of the temporal variations in comet K2, we targeted instances when the Earth crossed the projected orbital plane of the comet, in order to probe the out-of-plane extent of the coma free from the effects of projection.

A journal of observations is given in Table (\ref{geometry}), where dates are expressed as  Day of Year (DOY) with DOY = 1 on UT 2017 January 01.

\subsection{Morphology}

\begin{figure}
\epsscale{0.95}
\plotone{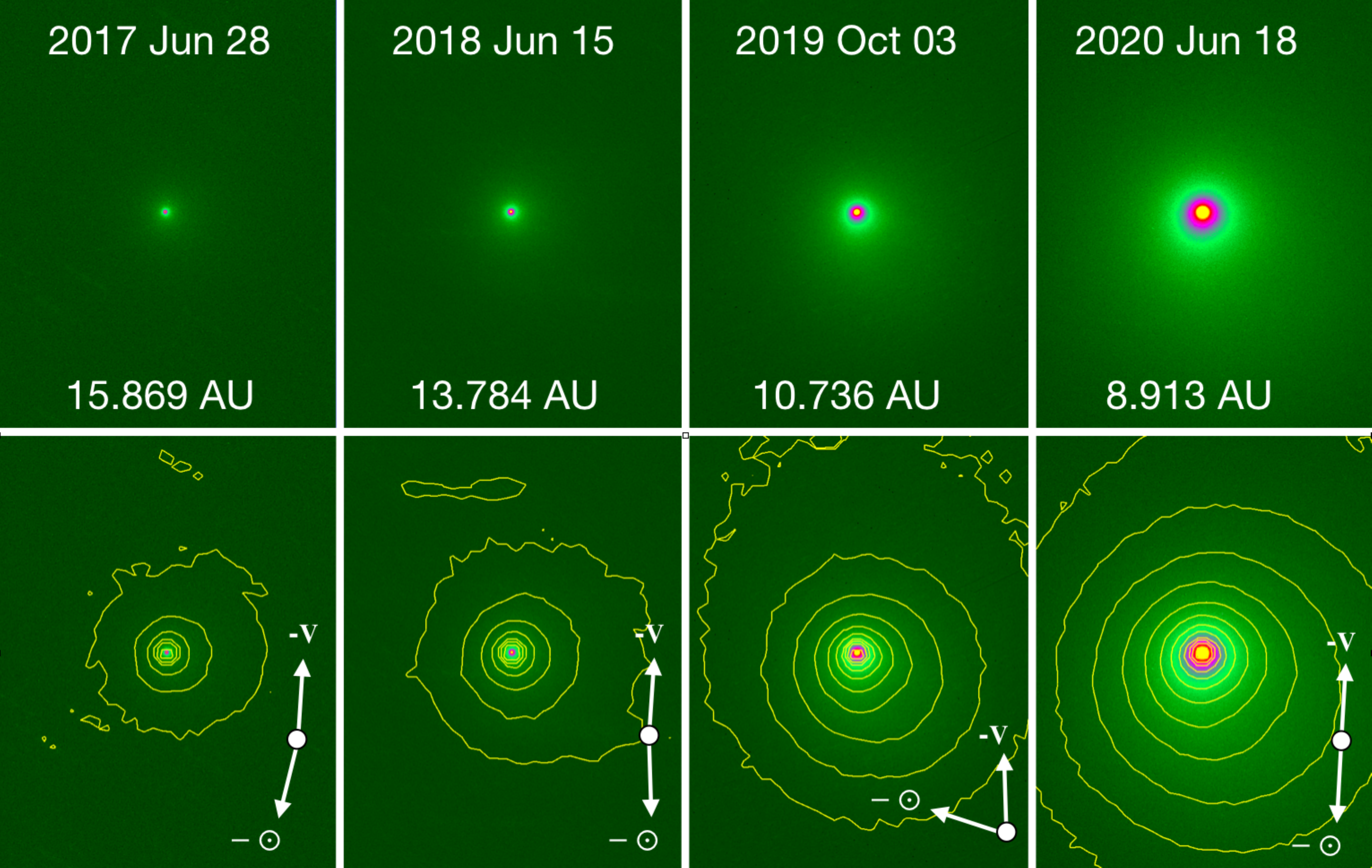}

\caption{Images of C/2017 K2 on four epochs shown in the top row with common scaling from -0.05 to 2.0 data numbers per second, showing the steady brightening of the comet.  The bottom row shows the same four images but with contours  spaced by a factor of two in surface brightness starting from 0.01.  Each panel has North to the top, East to the Left, and is 32\arcsec~tall.    The heliocentric distances of the comet are indicated, as are the projected directions of the anti-solar vector ($-\odot$) and the negative heliocentric velocity vector ($-V$). \label{images}}
\end{figure}

\begin{figure}
\epsscale{0.95}
\plotone{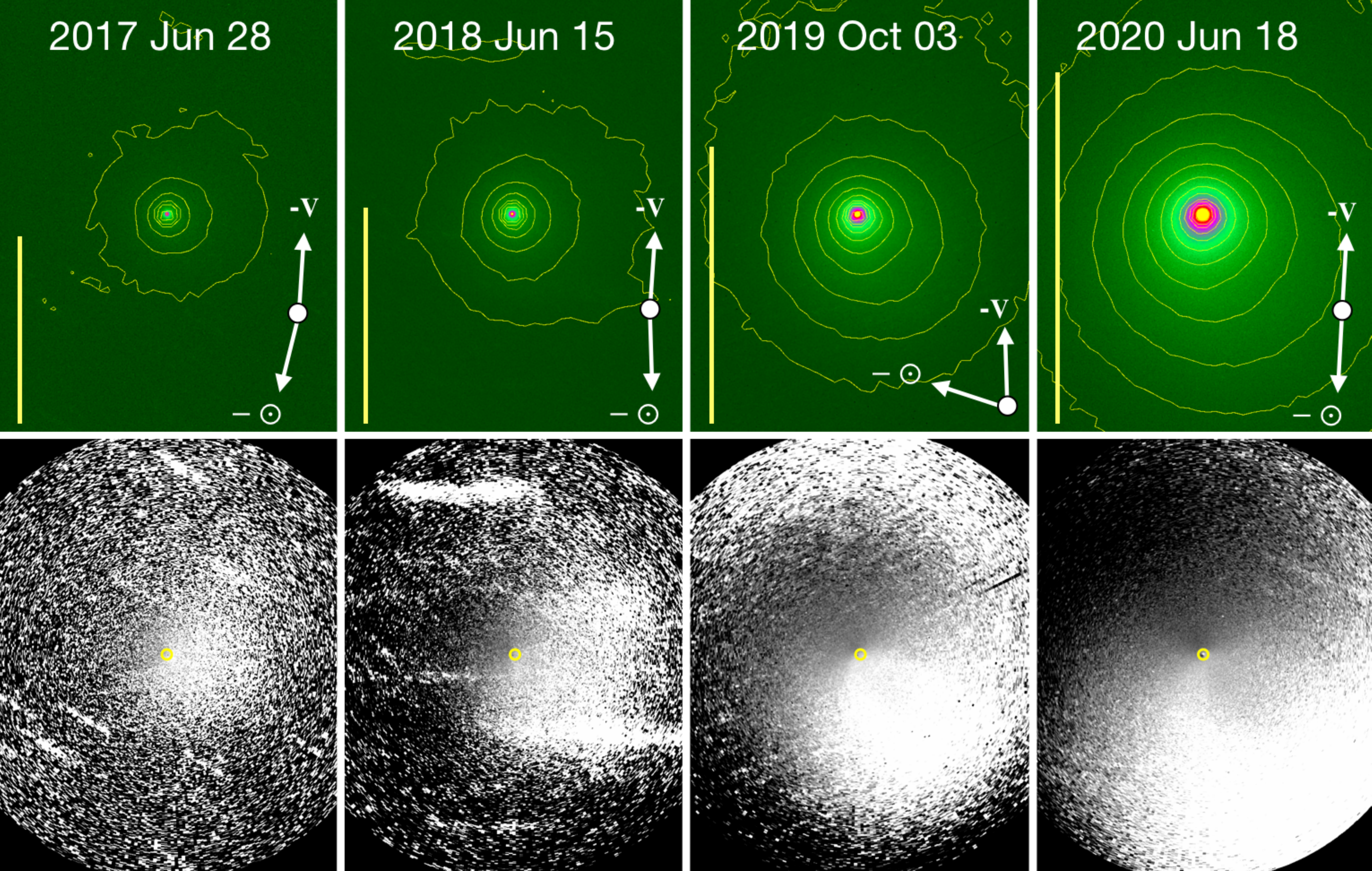}
\caption{(Top row:) Same as Figure \ref{images} to give the scale (vertical yellow bar is 160,000 km in length), orientation and direction arrows for comparison with (Bottom row:) Images spatially filtered by dividing each image by the annular median. The location of the nucleus is marked in each panel by a yellow circle. \label{rings}}
\end{figure}

The absence of a distinct tail  (Figure \ref{images}), as we previously noted (Jewitt et al~2017, 2019), reflects the minimal effect of solar radiation pressure on large ($a \ge$ 100 $\mu$m) particles.  Some deviations from circular symmetry of the isophotes are present, however, hinting at structure in the pattern of emission from the nucleus.   To enhance this azimuthal coma structure, we divided each image by the annular median, using the excellent software made available by Samarasinha and Larson (2014).  The annuli were taken to be 1 pixel (0.04\arcsec) wide and divided into 1\degr~sectors, each centered on the nucleus.  Figure (\ref{rings}) shows the results, further smoothed by convolution with a Gaussian function of 3 pixels (0.12\arcsec) half-width at half-maximum to reduce the noise.  Other than for excursions due to imperfectly removed field objects, the panels show excess fan-like coma projected to the south west of the nucleus, with a central axis of the fan rotating clockwise from position angle 250$\pm$20\degr~in 2017 to 170$\pm$10\degr~in 2020. The generally smooth time evolution of the fan orientation suggests that there is no direct link with the projected anti-solar direction, which changes appreciably over the period of observations (Figure \ref{rings}). This is likely a result of the extreme foreshortening resulting from the small phase angles at which K2 is observed.  Regardless of the orientation of K2's rotation axis, the nucleus should not have experienced strong seasonal changes between 2017 and 2020.

The observation on UT 2020 June 18 was taken only 0.08\degr~from the projected orbital plane of comet K2, proving that ejection is asymmetric, with more dust ejected to the south in the plane of the sky.

\subsection{Aperture Photometry}

We used photometry to measure the scattering cross-section of the coma.  The angular sizes of the photometry apertures were scaled to take account of the varying geocentric distance, $\Delta$, so as to maintain fixed linear radii of 5, 10, 20, 40, 80 and 160$\times$10$^3$ km when projected to  the distance of the comet.   The use of fixed linear (as opposed to angular) apertures guarantees that we obtain consistent measurements of a fixed volume around the nucleus. This removes the need to make an additional (and uncertain) geometry-dependent photometric corrections in order to meaningfully compare measurements taken at different times.  The smallest aperture (5000 km at the comet) was picked so as to remain larger in angular extent than the 0.4\arcsec~radius of the PSF of the telescope.  The largest aperture (160,000 km at the comet) was picked to match the field of view of WFC3.  The background sky brightness and its uncertainty were estimated from the median and dispersion of data numbers in a concentric annulus as large as permitted by the field of view of WFC3, typically corresponding to  inner and outer radii of 400 and 500 pixels (16\arcsec~and 20\arcsec), respectively.  The photometry was calibrated assuming that a G2V source with V = 0 would give a count rate 4.72$\times$10$^{10}$ s$^{-1}$ in the same filter\footnote{\url{http://etc.stsci.edu/etc/input/wfc3uvis/imaging/}}. 

The apparent V magnitudes were converted to the scattering cross-section of dust using

\begin{equation}
C =  \frac{1.5\times 10^6}{p_V} 10^{-0.4 H}
\label{area}
\end{equation}

\noindent where $p_V$ is the geometric albedo and $H$ is the absolute magnitude computed from

\begin{equation}
H  = V - 5\log_{10}(r_H \Delta) + 2.5\log_{10}\Phi(\alpha).
\label{H}
\end{equation}

\noindent Quantity $\Phi(\alpha) \le 1$ is the phase function, equal to the ratio of the brightness at phase angle $\alpha$ to that which would be observed at the same $r_H$ and $\Delta$ and $\alpha$ = 0\degr.  Neither $\Phi(\alpha)$ nor  $p_V$  is observationally constrained in comet K2 and so we are forced to make assumptions in order to make progress.  We assumed $2.5\log_{10}\Phi(\alpha) = -B \alpha$, with $B$ = 0.02 magnitudes degree$^{-1}$, consistent with observations of other active comets (c.f. Meech et al.~1986).  The largest phase angle in our data is $\alpha$ = 6\degr~(Table \ref{geometry}), so that the effects of even a factor-of-two error in $\Phi(\alpha)$ are unimportant.   However, we emphasize that we have no specific constraints on the phase function of K2 and the uncertainties could, in principle, be larger.  We assume $p_V$ = 0.1 in Equation (\ref{area}) in order to be consistent with results for other cometary dust comae (Zubko et al.~2017).    Results for other values of $p_V$ can be simply scaled from Equation \ref{area}. The apparent and absolute magnitudes as well as the derived cross-sections are listed in Table (\ref{photometry}).

\section{DISCUSSION}

\subsection{Heliocentric Index, Velocity Law and the Lag Effect}
\begin{figure}
\epsscale{0.8}
\plotone{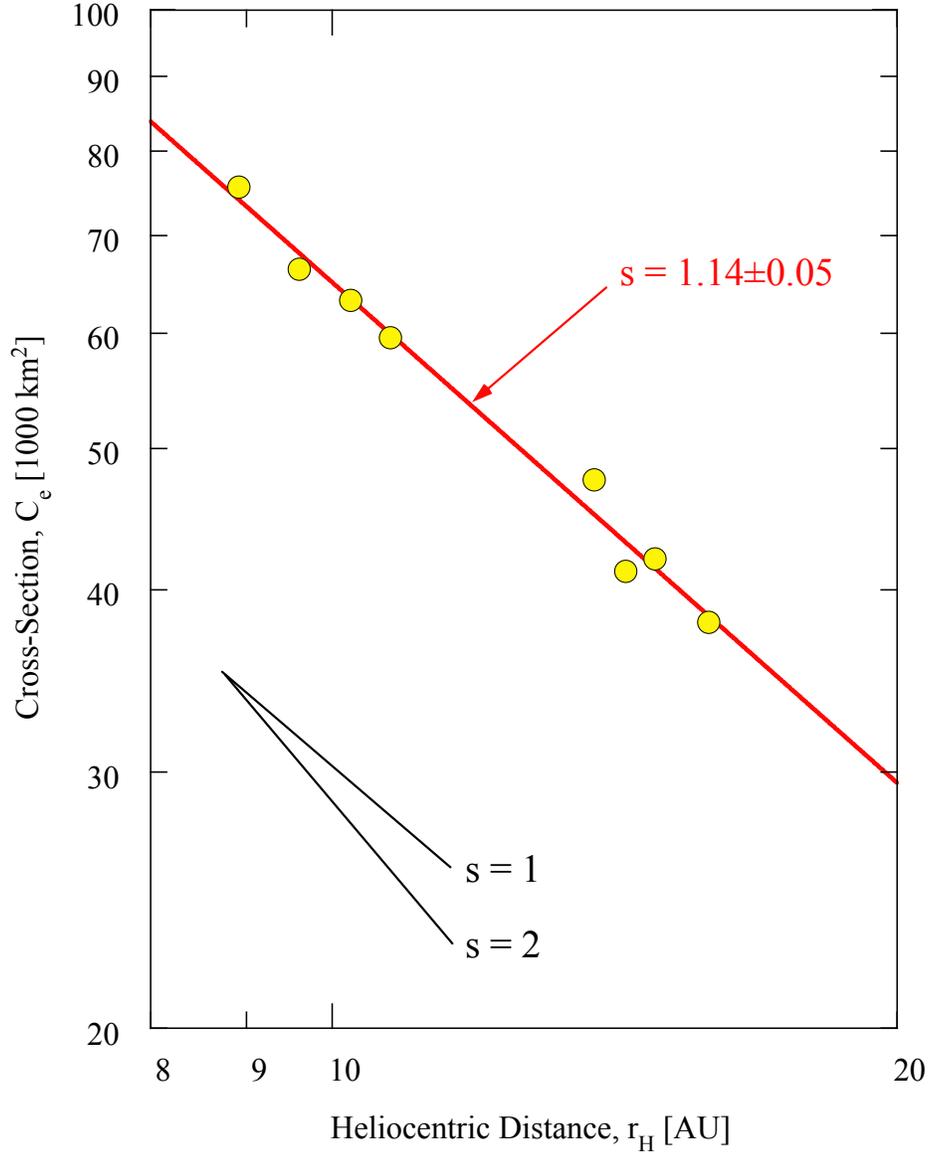}
\caption{Total scattering cross-section of K2 (yellow-filled circles) as a function of the heliocentric distance.    The cross-section was computed as described in the text from Equation (\ref{area}). Equation (\ref{fit}) is shown as a solid red line.     Black lines ilustrate heliocentric indices $s$ = 1 and 2, as labeled.  
\label{C160_vs_rH}}
\end{figure}

\textbf{Heliocentric Index:} Figure (\ref{C160_vs_rH}) shows the cross-section in the 160$\times 10^3$ km fixed-radius aperture as a function of heliocentric distance.  We  fitted a weighted power-law to the data by least-squares using

\begin{equation}
C = C_0 r_H^{-s} 
\label{fit}
\end{equation}

\noindent  where $C_0$ and $s$ are constants and $s$ is the ``heliocentric index''.  Quantity $C_0$ is the cross-section in the 160$\times 10^3$ km aperture scaled to $r_H$ = 1 AU using index $s$.   We find  $C_0 = (907\pm112)\times10^3$ km$^2$ and  heliocentric index $s = 1.14\pm0.05$ (solid red line in Figure  \ref{C160_vs_rH}).  For comparison, in equilibrium sublimation with sunlight, the production rate of an exposed supervolatile should vary as $r_H^{-2}$ (i.e.~$s$ = 2), which is steeper than the measured value by about 17$\sigma$. We attribute this difference to two effects; 1) the heliocentric distance dependence of the dust velocity and 2)  a ``lag effect'', both of which we describe below.   Note that we ignore the cross-section of the nucleus, $C_n$, in computing the heliocentric index.  The limiting nucleus radius $r_n <$ 9 km (Jewitt et al.~2019) gives $C_n < 250$ km$^2$, which is  $<$1\% of even the smallest cross-section (3.8$\times10^4$ km$^2$) measured within the 160$\times 10^3$ km photometry aperture (Table \ref{photometry}).

\textbf{Dust Velocity:} The gas drag force acting on a spherical particle of density $\rho_s$ and radius $a$ is $F = C_D \pi a^2 \rho_g(r,r_H) (V_g - V)^2$, where $C_D$ is a dimensionless drag coefficient of order unity, $\rho_g(r,r_H)$ is the density of the gas at distance $r$ from the nucleus and $r_H$ from the Sun, $V_g$ the bulk velocity of the gas and $V$ the speed of the particle.  Quantity $V_g - V$ is the speed of the gas relative to the particle, decreasing as the particle accelerates away from the nucleus.  We write 

\begin{equation}
\rho_g(r,r_H) = \rho_0 \left(\frac{r_n}{r}\right)^2 \left(\frac{10}{r_H}\right)^2
\label{density}
\end{equation} 

\noindent with  $r_n$ being the radius of the nucleus, $r_H$ the heliocentric distance in AU, and $\rho_0$  the gas density at the nucleus surface ($r = r_n$)  normalized to heliocentric distance $r_H = 10$ AU. The $r_H^{-2}$ dependence in Equation (\ref{density}) is appropriate for supervolatile ices, in which almost all energy absorbed from the Sun is used to break molecular bonds, leaving little to sustain thermal radiation (e.g.~Jewitt et al.~2017).  It is not appropriate for water ice, which shows a steeper distance dependence, except at heliocentric distances $r_H \lesssim$ 1 AU.  However, water ice is involatile at the heliocentric distances and temperatures considered here.

Substituting for $\rho_g(r,r_H)$ and neglecting the gravitational attraction to the nucleus for simplicity, the equation of motion is

\begin{equation}
\frac{4}{3}\pi\rho_s a^3 V \frac{dV}{dr} = C_D \pi a^2 \rho_0 \left(\frac{r_n}{r}\right)^2 \left(\frac{10}{r_H}\right)^2 (V_g - V)^2.
\label{EOM}
\end{equation}

\noindent The terminal speed of the particle, $V_{\infty}$, is given by integrating Equation (\ref{EOM}) as

\begin{equation}
\int_0^{V_{\infty}}\frac{V dV}{(V_g -V)^2} = \frac{3C_D}{4a} \frac{\rho_0}{\rho_s} \left(\frac{10}{r_H}\right)^2 r_n^2 \int_{r_n}^{\infty} \frac{dr}{r^2}.
\label{EOM2}
\end{equation}

\noindent Prompted by the low measured speeds of grains in K2, we assume $V \ll V_g$ and assume $V_g$ is independent of $r_H$, as found empirically by Biver et al.~(2002) in the distance range 7 to 14 AU, to solve Equation (\ref{EOM2})

\begin{equation}
V_{\infty} = \left(\frac{3C_D}{2} \frac{\rho_0}{\rho_s} \frac{r_n}{a}\right)^{1/2} \left(\frac{10}{r_H}\right) V_g.
\label{Vinfty}
\end{equation}

\noindent The inverse dependence of $V_{\infty}$ on $r_H$, for particles of a given size, contributes to the small value of the heliocentric index because the residence time in a given fixed-radius aperture decreases as the comet approaches the Sun, even as the production rate increases.  

To calculate $V_{\infty}$ from Equation (\ref{Vinfty}), we substitute $\rho_0 = f_s/V_g$, where $f_s$ (kg m$^{-2}$ s$^{-1}$) is the equilibrium mass sublimation flux evaluated, by definition, at heliocentric distance $r_H$ = 10 AU.  The hemispheric sublimation rate is given, to first order, by equating the  power absorbed from the Sun to the power consumed in sublimation,

\begin{equation}
f_s = \frac{L_{\odot}}{8\pi r_H^2 \mathcal{H}},
\end{equation}

\noindent where $L_{\odot} = 4\times10^{26}$ W is the luminosity of the Sun, $r_H$ is the heliocentric distance expressed in meters, and $\mathcal{H} = 2\times10^5$ J kg$^{-1}$ is the latent heat of vaporization of CO (Huebner et al.~2006).  Substituting, we find $f_s \sim 3\times10^{-5}$ kg m$^{-2}$ s$^{-1}$ at 10 AU.  A more detailed (numerical) solution of the energy balance equation, including a term for radiation cooling of the ice, gives a slightly smaller $f_s = 2\times10^{-5}$ kg m$^{-2}$ s$^{-1}$.  (We note that this number is, itself, a probable over-estimate of $f_s$ given that  sublimation likely proceeds from beneath a thin, porous mantle, not from the exposed surface).   The drag coefficient, $C_D$, is unknown.  For simplicity, we set $3C_d/2$ = 1, which should be good to within a factor of order 2. Then, with $V_g$ = 130 m s$^{-1}$, $\rho_s$ = 500 kg m$^{-3}$ and $r_n \le 9$ km (Jewitt et al.~2017), we obtain, for a radius $a$ = 1 mm particle,  $V_{\infty} \le  4.6$ m s$^{-1}$  at $r_H$ = 14.8 AU (the average distance of K2 used to estimate the speed in Jewitt et al.~2019).  This is acceptably close to the measured speed, $V_{\infty} \sim$ 4 m s$^{-1}$, given the simplicity of the model and the fact that many of the quantities in Equation (\ref{Vinfty}) are unmeasured.  

\begin{figure}
\epsscale{0.8}
\plotone{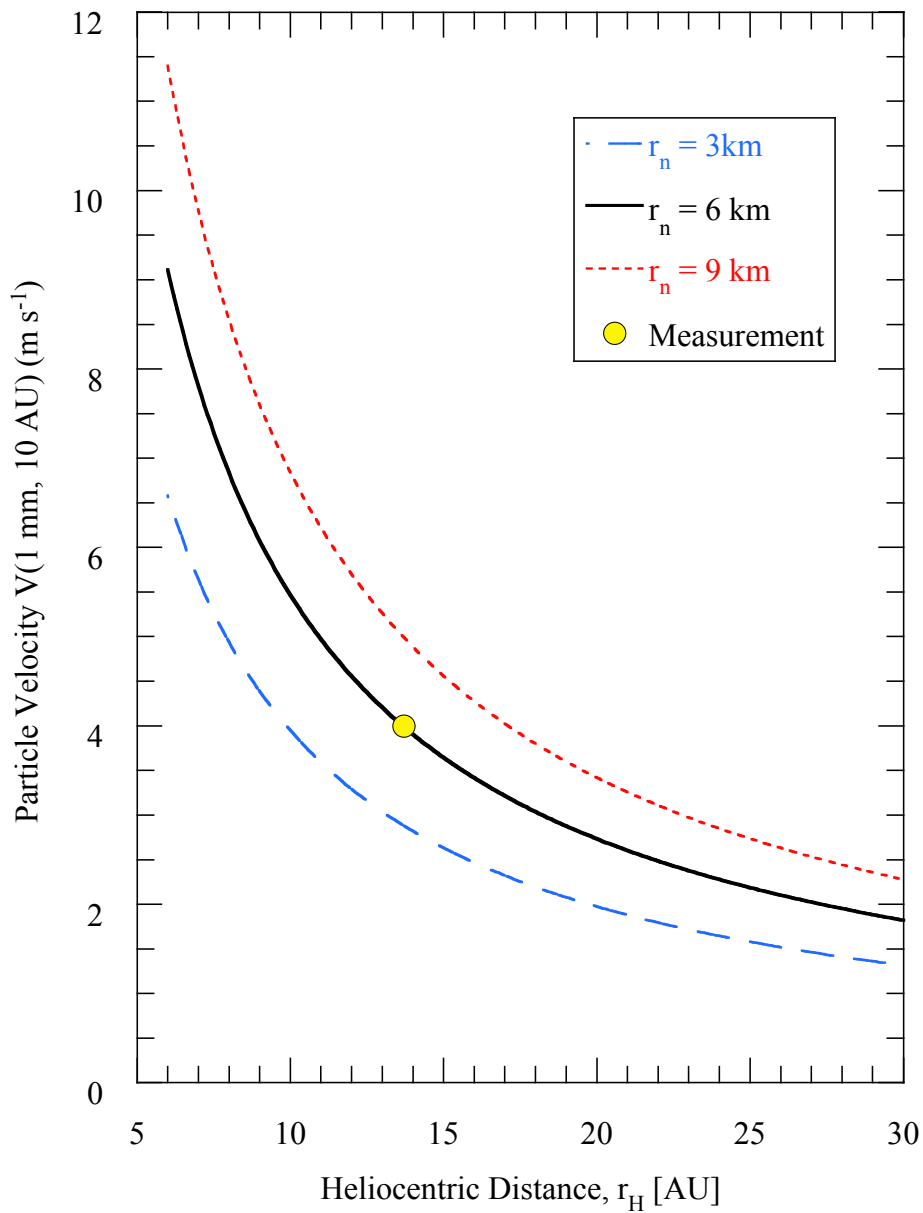}
\caption{Velocity of 1 mm radius particles as a function of heliocentric distance.  The  curves were computed from Equation (\ref{Vinfty}) using $r_n$ = 3 km (long-dashed blue line), $r_n$ = 6 km (solid black line) and $r_n$ = 9 km (short-dashed red line), respectively.  The yellow-filled circle shows the speed measured using a Monte Carlo simulation of  data on UT 2018 June 15. 
\label{velocity_plot}}
\end{figure}

Equation (\ref{Vinfty}) can be written
\begin{equation}
V_{\infty} = 2.3 \left(\frac{r_n}{1~\textrm{km}}\right)^{1/2} \left(\frac{1~\textrm{mm}}{a}\right)^{1/2} \left(\frac{10~\textrm{AU}} {r_H}\right)
\label{Vinftynumbers}
\end{equation}

\noindent with $a$ in millimeters, $r_n$ in km and $r_H$ in AU.  This relation is plotted in Figure (\ref{velocity_plot}) for a 1 mm particle and nucleus radii $r_n$ = 3 km (long-dashed blue curve), $r_n$ = 6 km (solid black curve) and $r_n$ = 9 km (short-dashed red curve).  The particle speed reported by Jewitt et al.~(2019), marked for comparison in the figure, is in reasonable agreement with Equation (\ref{Vinftynumbers}).


%
\textbf{Lag Effect:} At a given epoch, $C$ is contributed by particles having a wide range of sizes, ejection speeds and aperture residence times, some of which are very long.   Under gas drag acceleration, small, fast-moving particles quickly leave the photometry aperture, but large slowly-moving particles  linger  longer.   In distant comet K2, the residence times can be extreme.  For example,  1 mm  particles have characteristic ejection speeds $V =$ 4 m s$^{-1}$ (Jewitt et al.~2019) at which speed they  take $\sim 4\times10^7$ s (1.3 years) to travel across the radius of the 160,000 km aperture. Larger, slower particles  would taken even longer.  Particles released into the coma at any instant thus contribute to a ``background'' of lingering particles accumulated over the previous year or, depending on their size and ejection speed, even longer.   

To explore this lag effect with a more physical model, we constructed a set of Monte Carlo simulations. The simulations follow the motions of particles  ejected  following the velocity law in Equation (\ref{Vinftynumbers}) and starting at heliocentric distance $r^{\star}$.  We   assumed that the heliocentric dependence of the gas production rate varies as $dM/dt \propto r_H^{-2}$, needed to maintain equilibrium with the  insolation, and we included particles with dimensionless radiation pressure parameters in the range $\beta = 10^{-3}$ to $10^{-5}$.  These  correspond to particle radii in the range 1 mm to 100 mm, respectively.


The distribution of particle radii is taken to be a power-law, such that the number of particles having radii in the range $a$ to $a+da$ is $n(a) da = \Gamma a^{-q}da$, with $\Gamma$ and $q$ being constant.  In Table (\ref{indices}) we list a brief and incomplete summary of values of $q$ reported in the recent literature.  Most are based on matching the isophotal dust distribution in optical images but for three comets, 1P/Halley, 67P/Churyumov-Gerasimenko and 103P/Hartley, we also list direct  measurements obtained from spacecraft in, or passing through, the coma.   There is a tendency for the size distribution to become steeper as the particle size increases.   The unweighted mean of the measurements in the Table is $q$ = 3.7$\pm$0.2 and the median is $q \sim$ 3.4.  We also attempted to form the weighted mean of the data by assuming that, where no error is quoted, the error is $\pm$0.2 and, where a range is quoted, the effective $q$ is the middle value and the error is half the range.  The resulting weighted mean is $q$ = 3.45$\pm$0.04.  Consistent with these determinations in other comets, we successfully represented the coma morphology of comet K2  by Monte Carlo models having $q$ = 3.5 (Jewitt et al.~2017, Hui et al.~2018).  Accordingly, we proceed to interpret the heliocentric index using models with $q$ = 3.5.

\begin{figure}
\epsscale{0.8}
\plotone{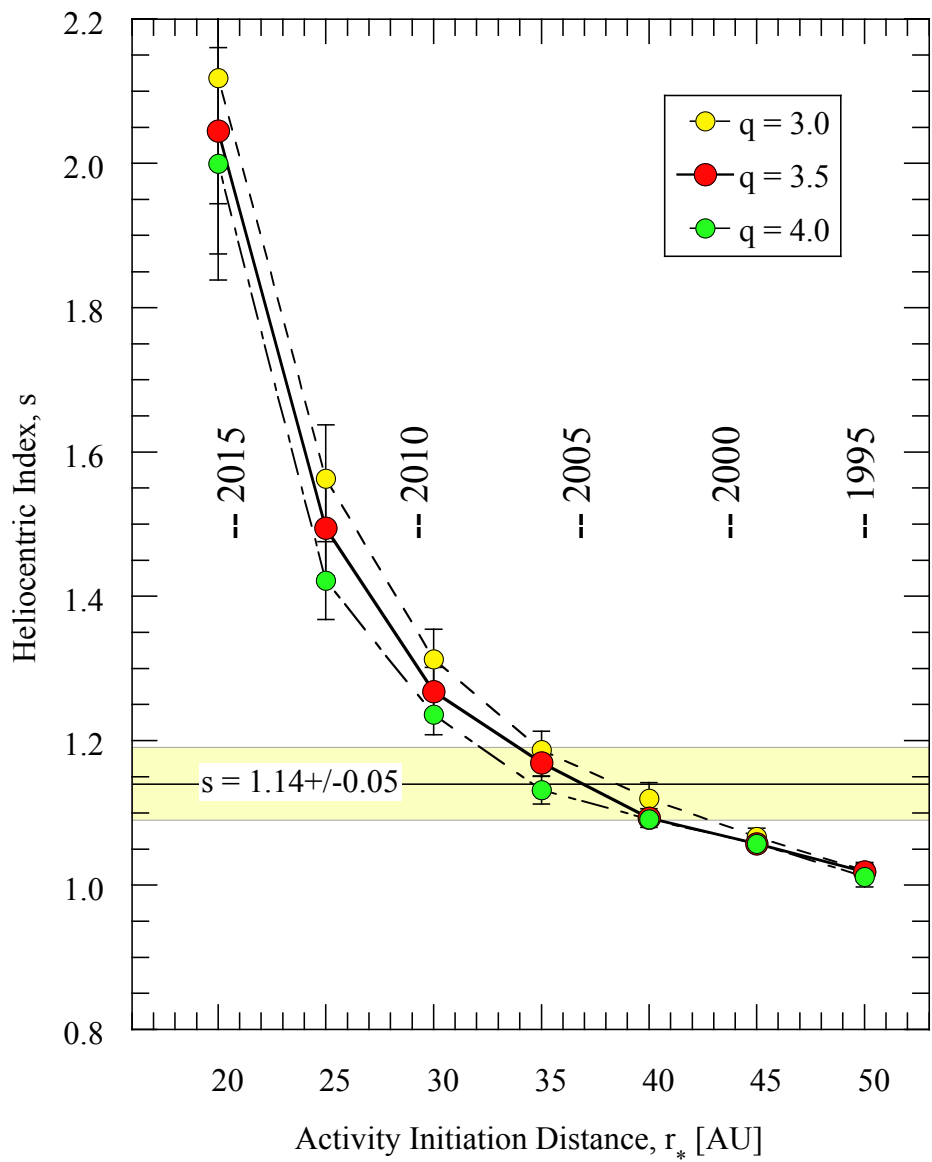}
\caption{Heliocentric index vs.~activity initiation distance.  Monte Carlo simulations show the model index for size distribution indices $q$ = 3.0 (yellow circles), 3.5 (red circles) and  $q$ = 4.0 (green  circles), respectively.  The measured heliocentric index and its $\pm1\sigma$ uncertainties are indicated in the yellow-shaded horizontal box.  Dates in the middle part of the figure show the heliocentric distances reached by  C/2017 K2 on approach to perihelion.
\label{mc_vs_rstart}}
\end{figure}

Results are shown in Figure (\ref{mc_vs_rstart}), which shows the modeled heliocentric index, $s$, as a function of $r^{\star}$.  Error bars on the model points were calculated from least-squares fits at five heliocentric distances spanning the 9 AU to 16 AU heliocentric distance range, to simulate the way in which we computed $s$ from the HST data.  Also marked in the figure are the measured value, $s = 1.14\pm0.05$, and a shaded region extending $\pm1\sigma$ from this value.      
We show models for three values of the size distribution index, $q$.  

For the nominal $q = 3.5$ size distribution index, we deduce an initiation distance $r_{\star}$ = 36 AU.  Adopting $q$ = 3.0, a value smaller than typical of comets (Table \ref{indices})  the minimum turn-on distance is $r^{\star} \ge$ 38 AU (c.f.~yellow circles in Figure \ref{mc_vs_rstart}; reached by K2 in 2004).     Index values $q = 4.0$ result in initiation distances $r_{\star} >$ 34 AU.  Model values of $s$ within $\pm$1$\sigma$ of the measured value are obtained for 32 $\le r_{\star} \le 42$ AU for 3.0 $\le q \le $ 4.0.  Overall, we conclude that the sensitivity of $r_{\star}$ to the adopted size distribution is modest unless  $q$  is pathologically large or small.  The latter possibility is counter-indicated by our own Monte Carlo models of the optical data in which $q$ = 3.5 provides a convincing match to the morphology (Jewitt et al.~2017, Hui et al.~2018).  In short, the shallow heliocentric index requires that activity in K2 begin at Kuiper belt distances.

This lag effect on the photometry is particularly prominent in K2 because of the unusually large  distances at which the comet has been observed, because of the large average  size (and low speed)  of the particles, and because of the head-on viewing geometry,  resulting in very long aperture residence times. Published measurements of other comets are difficult to compare with the present study because, for example, no other studies have sampled in-bound comets over heliocentric distances as  large  as those considered here.   Furthermore, published photometry generally uses fixed angular (not linear) apertures, requiring an uncertain correction for the changing volume of coma that is measured as $r_H$ varies.  Sekanina's (1973) early inference  that long-period comets Baade 1955 VI and Haro-Chivara 1956 ejected icy sub-millimeter particles when at 5 to 15 AU, while lacking photometric support and being less extreme than the case of comet K2, is perhaps the most similar to the picture  developed here.

\subsection{Size of the Coma}
\begin{figure}
\epsscale{0.95}
\plotone{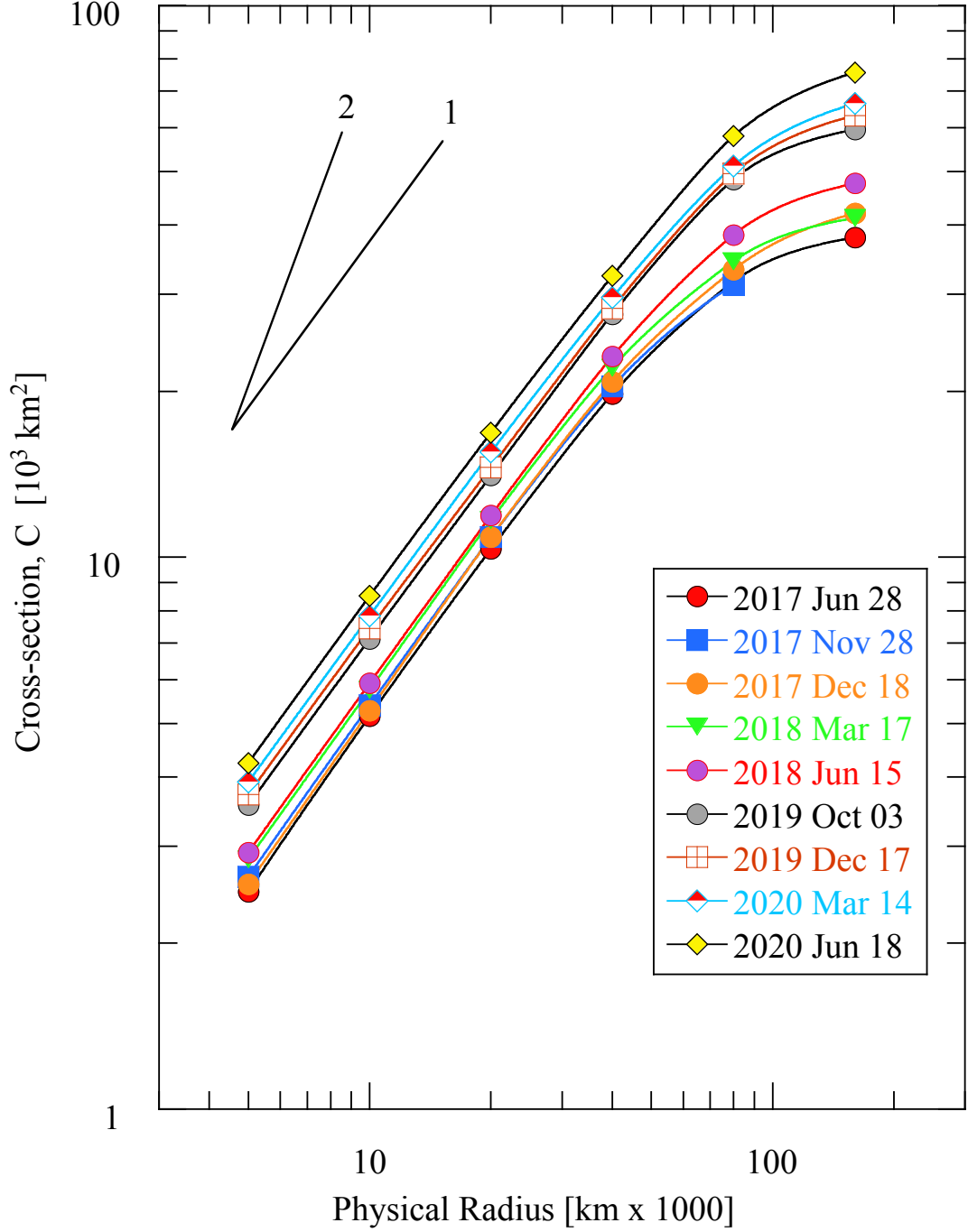}
\caption{Cumulative dust cross-section as a function of aperture radius for each date of observation. Power law slopes in $C \propto r^x$, with $x$ = 1 and 2 are marked for comparison with the data.
\label{SB_plot}}
\end{figure}

The radial distribution of dust within the coma was assessed using the annular photometry from Table (\ref{photometry}).  Figure (\ref{SB_plot}) shows the cumulative dust cross-section as a function of the linear aperture radius, for each of the dates of observation in the table. The cumulative profiles are similar in shape on each date, consistent with the steady appearance of the comet, but show progressive increases in brightness at all radii between 2017 and 2019.  In steady-state, the surface brightness  of a coma varies inversely with radius, $r$, because of the equation of continuity (Jewitt and Meech 1987).  When integrated with respect to radius, as in Figure (\ref{SB_plot}), the encircled brightness  of a steady-state coma should vary in proportion  to $r$.  This accurately describes the coma of K2 for all measured profiles up to radii $\sim$80,000 km, beyond which the cumulative profile flattens.  This truncated profile shape is typical of comets, where the cause is either the imposition of a sunward ``nose'' to the coma through the action of radiation pressure (Jewitt and Meech 1987), or the existence of ``fading grains'', which darken or disintegrate in response to space exposure (Baum et al.~1992).

\begin{figure}
\epsscale{0.95}
\plotone{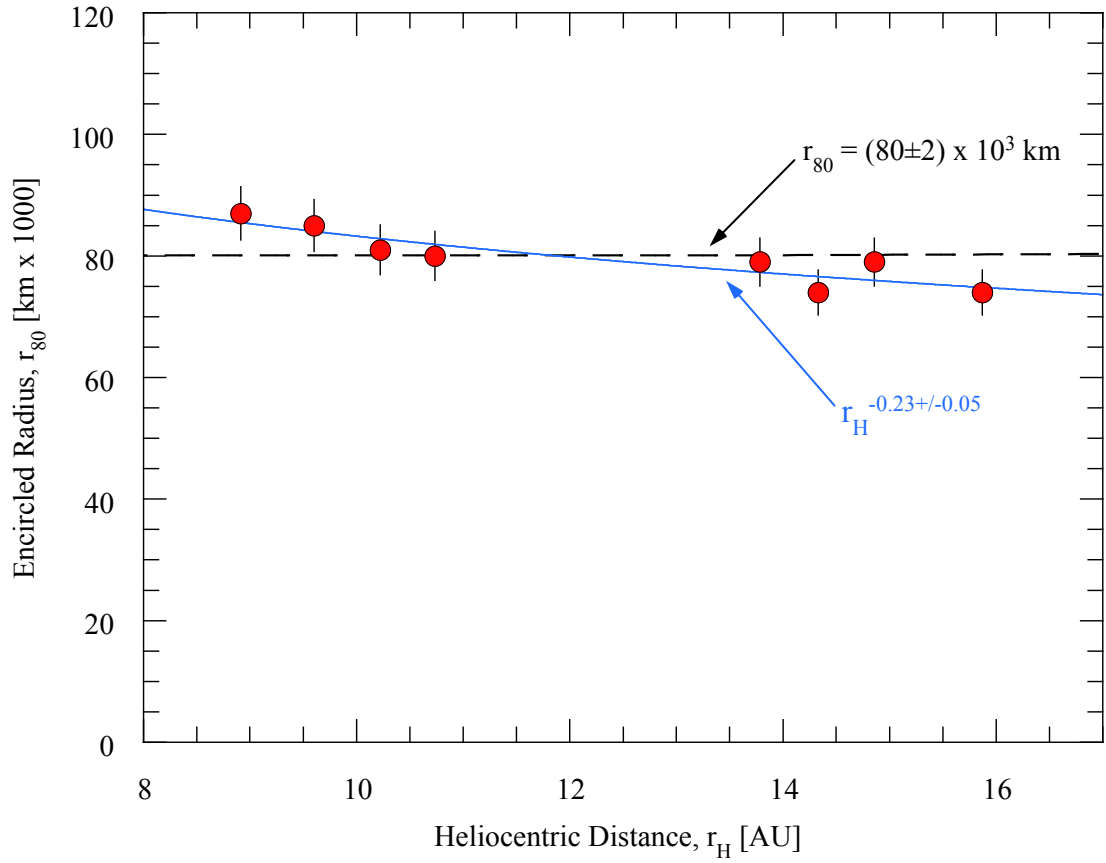}
\caption{Coma radius $r_{80}$ as a function of heliocentric distance, $r_H$. 
\label{r80_plot}}
\end{figure}

We define  the effective radius of the coma by $r_{80}$, the linear radius at which the encircled cross-section is 80\% of the peak cross-section  determined using the 160,000 km radius aperture.  Measurements of $r_{80}$ are plotted as a function of the heliocentric distance in Figure (\ref{r80_plot}).  The average value is $r_{80} = (8.0\pm0.2)\times10^7$ m (error on the mean of eight measurements).  Travel times for coma particles, estimated using  $\tau = r_{80}/V$ are $\sim$1 year, given the  $V \sim$4 m s$^{-1}$ speeds of 1 mm particles (Jewitt et al.~2019).  Figure (\ref{r80_plot}) shows evidence for a weak gradient.  A least-squares fitted power law gives $r_{80} = (142\pm16)\times10^3 r_H^{-0.23\pm0.05}$ in the range 9 $\lesssim r_H \lesssim$ 16 AU.
The near constancy of $r_{80}$ argues for the action of radiation pressure and against the ``fading grains'' hypothesis, as we argue below.

A dust particle launched sunward at speed $V$ will be stopped by radiation pressure at a turn-around distance, $\ell$, given by $\ell  = V^2/(2 \beta g_{\odot})$.  Here, $g_{\odot}$ is the gravitational acceleration towards the Sun and $\beta$ is the dimensionless radiation pressure efficiency factor, such that $\beta g_{\odot}$ is the acceleration of the particle.  Quantity $\beta$ is inversely related to particle size (Bohren and Huffman 1983) and conveniently approximated by $\beta \sim 10^{-6}/a$, with $a$ expressed in meters.  A 1 mm particle has $\beta \sim 10^{-3}$.  The solar gravity may be written $g_{\odot} = g_{\odot}(1)/r_H^2$, where $g_{\odot}(1)$ = 0.006 m s$^{-2}$ is the acceleration at 1 AU and $r_H$ is in AU.  Then, setting $V = V_{\infty}$ and with the use of  Equation (\ref{Vinfty}), we find

%

\begin{equation}
\ell = 100 \left(\frac{3C_D}{2} \frac{\rho_0}{\rho_s} \frac{r_n}{a}\right)  \frac{V_g^2 }{2\beta g_{\odot}(1)}
\label{ell}
\end{equation}

\noindent which is independent of heliocentric distance.

Substituting into Equation (\ref{ell})  for the parameters as above, we obtain $\ell \sim 10^8$ m, in good agreement with the measured $r_{80} = 8\times10^7$ m.  The observation that $\ell \sim r_{80}$ and the fact that both quantities are approximately independent of $r_H$ strongly favor radiation pressure shaping of the coma as the cause of the flattened profiles in Figure (\ref{SB_plot}).

The ``fading grains'' hypothesis is less consistent with the data.   In its favor is the fact that the residence times for particles in the coma are very long, $\sim$ 1 year, allowing time for space weathering to act, or for weakly bonded aggregate particles to disaggregate.  However, weathering and disaggregation should become more pronounced as K2 approaches the Sun, leading to a coma radius that shrinks as $r_H$ decreases.  This is because the insolation varies as $r_H^{-2}$ while the residence time in the coma varies, by Equation (\ref{Vinfty}), as $r_H^{-1}$.  Therefore, weathering and disaggregation should be more pronounced at smaller $r_H$, leading to shrinkage of the coma upon approach to the Sun.   Since this is not observed (Figure \ref{SB_plot}), we discount the fading grains hypothesis.

\subsection{Mass Loss Rate}
 The mass of an opaque  spherical particle, $M$, is proportional to its geometric cross-section, $C$, according to $M = 4\rho_s a C/3$, where $\rho_s$ is the particle density and $a$ is the  particle radius.   An equivalent relation holds for an optically thin collection of spheres, with $a$ replaced by the mean particle radius $\overline{a}$, and $C$ being the sum of the cross-sections of all the particles within the projected photometry aperture.  Since $C$ is measurable from the photometry using Equations (\ref{area}) and (\ref{H}), we can estimate the coma mass in K2.   At any instant, the coma mass within an aperture is
 
 \begin{equation}
 M(r_H) = (4/3)\rho_s \overline{a} C(r_H).
 \label{mass}
 \end{equation}

Dust within an aperture of radius $r$ must, in steady state, be replaced on the crossing timescale $\tau = r/V$.  Differentiating Equation (\ref{mass}), setting $dC(r_H)/dt = C(r_H)/\tau$ and neglecting the numerical multiplier, we estimate the steady-state dust loss rate from

\begin{equation}
\frac{dM}{dt} = \rho_s \overline{a}\frac{C(r_H) V(r_H)}{r}
\end{equation}

\noindent     Substituting for $V$ from Equation (\ref{Vinftynumbers}), we have

\begin{equation}
\frac{dM}{dt} = \frac{2.3\times10^{-3} \rho_s } {r} \left(\frac{r_n}{1\textrm{km}}\right)^{1/2} \left(\frac{ \overline{a}}{1 \textrm{mm}}\right)^{1/2}  \left(\frac{10} {r_H}\right) C(r_H)
\label{dmbdt}
\end{equation}

\noindent The rate given by Equation (\ref{dmbdt}) is only an order-of-magnitude estimate of the mass loss rate in dust because of the many unmeasured parameters.  For example, the particle density, $\rho_s$, is unmeasured, we possess only an upper limit to the nucleus radius,  $r_n \le$9 km, the mean particle radius, $\overline{a}$, is uncertain to within a factor of at least  two, and because the derived cross-sections, $C(r_H)$, rely on the assumption of the coma albedo, which itself could  be in error by a factor of two to three.  Nevertheless, the equation gives a useful measure of the relative mass production rates in dust and their variation with heliocentric distance.  

\begin{figure}
\epsscale{0.8}
\plotone{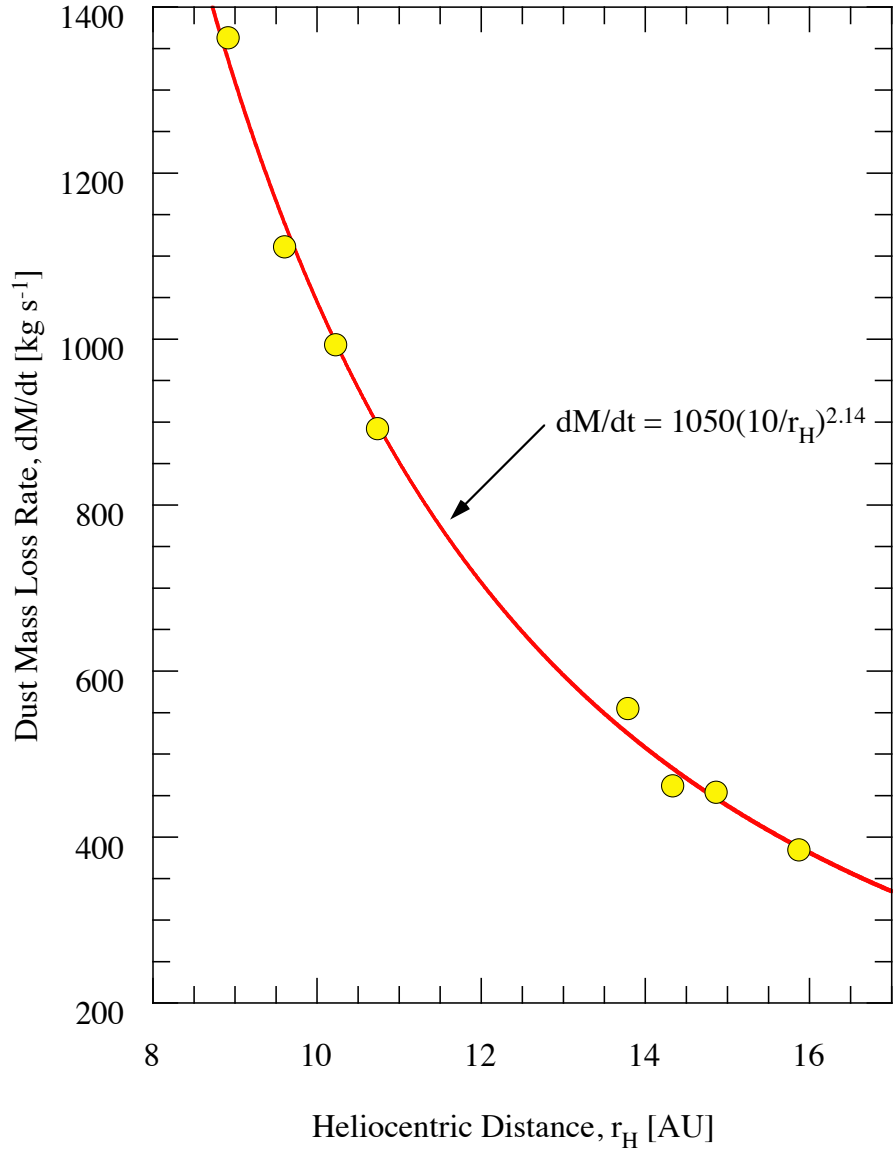}
\caption{Dust mass loss rate computed from Equation (\ref{dmbdt}) as a function of heliocentric distance.
\label{dmbdt_plot}}
\end{figure}

We plot Equation (\ref{dmbdt}) in Figure (\ref{dmbdt_plot}) assuming $r_n$ = 5 km, $\overline{a}$ = 1 mm, $\rho_s$ = 500 kg m$^{-3}$.  We find $dM/dt \sim 1050 (10/r_H)^{2.14}$ with $dM/dt$ in kg s$^{-1}$, over the 9 $\le r_H \le 16$ AU range.   At the upper end of this distance range, the derived mass loss rate is about a factor of two larger than obtained by Jewitt et al.~(2019) using less complete data and slightly different assumptions about the particle properties.  We regard the difference as insignificant.   Mass loss rates in K2 at 10 AU are comparable to those found in many comets at 1 AU, indicating the large mean particle size and the high level of activity in K2.


\subsection{Activity Mechanisms}

Dust mass loss rates implied by Equation (\ref{dmbdt}) can be supplied by freely sublimating  exposed CO of area $A \sim (dM/dt)/(f_s f_{dg})$, where $f_{dg}$ is the dust to gas ratio.  For example, dust production at rate $dM/dt$ = 400 kg s$^{-1}$ at 15 AU, where $f_s = 1\times10^{-5}$ kg m$^{-2}$ s$^{-1}$, requires a CO patch of area $A \sim 40 f_{dg}^{-1}$ km$^2$.  This sets a lower limit to the radius of a spherical nucleus $r_n \ge (A/\pi)^{1/2} \sim 3 f_{dg}^{-1/2}$ km.  Quantity $f_{dg}$ is unmeasured in comet K2, but in short-period comets, values $f_{dg} \ge$ 1 are normal. Extraordinarily large values (10 $\le f_{dg} \le$ 30) have been reported in comet 2P/Encke (Reach et al.~2000), where the large, mass-dominant  particles are reminiscent of those in K2.  If $f_{dg}$ = 20, for instance, the CO could be supplied from a sublimating area $A \sim$ 2 km$^2$ and a nucleus of radius $r_n \ge$ 0.7  km.  If, as is quite likely, sublimation proceeds from beneath a protective mantle of refractory material, $f_s$ will be over-estimated in this calculation and a larger sublimating area would be needed.

Gas drag forces resulting from the free sublimation of exposed supervolatiles would be sufficient to eject micron-sized grains against the gravity of the nucleus, but incapable of overcoming grain-grain cohesive forces (Gundlach et al.~2015, Jewitt et al.~2019) at these distances.  Paradoxically,  gas drag forces can overcome the (weaker) cohesive forces binding 1 mm sized particles but cannot eject them against the gravitational attraction to the nucleus because they are too heavy.  Therefore, taken at face value, no particles of any size can be ejected by gas drag, creating the so-called ``cohesion bottleneck" which operates beyond a critical distance that is controlled by the latent heat of sublimation of the responsible volatile and by cohesion.  One solution to the bottleneck problem might be  pressure build-up inside a porous medium having significant tensile strength, for example beneath a mantle or within postulated centimeter sized pebbles as proposed by Fulle et al.~(2020).   

Activity driven by the (exothermic) crystallization of amorphous water ice, leading to the release of trapped molecules (Prialnik et al.~2004), is possible at distances $r_H \lesssim$ 10 AU but not at the low equilibrium temperatures found at larger distances (Guilbert-Lepoutre 2012).   Comet K2 so far shows no evidence for excess activity that might be attributed to crystallization down to $r_H$ = 9 AU (Figure \ref{C160_vs_rH}), presumably indicating that amorphous ice is not present in close thermal contact with the surface.  However, the interpretation of this observation is ambiguous.  Amorphous ice could be absent in the nucleus, or it could simply have migrated to greater depths during a previous approach to the Sun.  In the latter case, crystallization might  play a future role, detectable by a surge in activity, as K2 approaches perihelion and the surface thermal wave diffuses into the interior.
A variety of non-thermal processes might also operate at large distances, including thermal fracture and electrostatic supercharging (Jewitt et al.~2019) but these are probably minor contributors to the activity.

\subsection{Surface Evolution}

Mass loss at rate $dM/dt$ corresponds to global erosion of a spherical nucleus of radius $r_n$ at the rate $dr_n/dt  = -(dM/dt)/(4\pi r_n^2 \rho$).  At $r_H$ = 15 AU, for example, $dM/dt \sim$ 400 kg s$^{-1}$ (Figure \ref{dmbdt_plot}) and a 5 km radius nucleus would shrink at the rate $dr_n/dt \sim$ -8 cm yr$^{-1}$.  At $r_H$ = 10 AU, $dM/dt \sim$ 1000 kg s$^{-1}$ and $dr_n/dt \sim -20$ cm yr$^{-1}$.   By extrapolation of Equation (\ref{dmbdt}), a meter or more of surface will be lost by the time K2 reaches $r_H$ = 5 AU, where water ice sublimation is expected to begin.  As on other comets, instead of being global, mass loss from K2 is likely to be confined to a fraction of the surface of the nucleus, with material lost locally from much smaller areas and greater depths than indicated by these global average values.  We conclude that topography on the nuclei of K2 and other long-period comets, even those entering the planetary region for the first time, can be substantially altered long before reaching the   water ice sublimation zone.

Cometary mass loss in the outer solar system may also account for one observational puzzle concerning the comets.  Specifically, numerous observations show that the optical colors of cometary nuclei and cometary dust are independent of the dynamical classification of the comet. Short-period and long-period comets are indistinguishable by their optical colors (Jewitt 2015).  For example, K2 has B-V = 0.74$\pm$0.02 and V-R = 0.45$\pm$0.02 (Hui et al.~2018), consistent with the mean colors of active short-period (B-V = 0.75$\pm$0.02, V-R = 0.47$\pm$0.02) and long-period (B-V = 0.78$\pm$0.02, V-R = 0.47$\pm$0.02) comets alike (Jewitt 2015).  It might be expected that the surfaces of long-period comet nuclei, which have been exposed  the galactic cosmic ray flux for billions of years, might have been chemically altered (Cooper et al.~1998) relative to the nuclei of short period comets, on which the surfaces are recently emplaced through the fall-back of sub-orbital debris (e.g. Marschall et al.~2020).  Distant activity in long-period comets may explain the similarity of colors by resurfacing the nuclei in fallback debris just as happens on the short-period nuclei.    Likewise, dust ejected into the comae of long period comets may come from beneath the surface layer chemically processed by cosmic rays.  

\subsection{Non-Gravitational Acceleration}
The recoil from anisotropic mass loss produces the non-gravitational acceleration of cometary nuclei.  To estimate the magnitude of the non-gravitational acceleration on the nucleus of comet K2, we assume that the ultimate source of the outflow momentum is the  expansion of gas, produced by sublimation at the nucleus surface, into the surrounding vacuum.  Dust is dragged from the nucleus by  the gas, which escapes at  speed, $V_{g}$.  While the momentum is always dominated by the gas,  the mass loss rate in more slowly moving dust can rival or exceed the mass loss rate in gas, giving rise to the ratio of dust to gas mass production rates  $f_{dg} > 1$.    Then,  the gas mass loss rate is $f_{dg}^{-1} (dM/dt)$ and the force exerted by the gas is $k_R (dM/dt) V_{g} /f_{dg}$.  The  momentum transfer coefficient, $k_R$, is equal to the fraction of the outflow momentum transferred to the acceleration of the nucleus, with  $k_R$ = 1 for collimated ejection and $k_R$ = 0 for isotropic emission.

 

The resulting magnitude of the non-gravitational acceleration is 

\begin{equation}
\alpha_{ng} = \frac{k_R V_{g}}{f_{dg}  M_n}\left(\frac{dM}{dt}\right)
\label{alphang}
\end{equation}

\noindent where  $M_n = (4\pi/3)\rho r_n^3$ is the mass of the nucleus, assumed to be spherical and of density $\rho$, radius $r_n$.   

Dynamically, what matters is the ratio of $\alpha_{ng}$ to the local acceleration due to the gravity of the Sun, $g_{\odot}$.  We define $\alpha' = \alpha_{ng}/g_{\odot}$ and write $g_{\odot} = G M_{\odot}/r_H^2$, where $G = 6.67\times10^{-11}$ N kg$^{-2}$ m$^2$ is the gravitational constant, $M_{\odot} = 2\times10^{30}$ kg is the mass of the Sun and $r_H$ is expressed in meters.   Then, we substitute for $M_n$ to find

\begin{equation}
\alpha' = \frac{3 k_R V_{g}}{4\pi G M_{\odot} \rho f_{dg} }  \left(\frac{\textrm{1 km}}{r_n}\right)^3 \left(\frac{r_H}{10\textrm{AU}}\right)^2  \left(\frac{dM}{dt}\right)
\label{accn}
\end{equation}

\noindent where $r_n$ is in kilometers, $r_H$ is in AU,  $dM/dt$ is in kg s$^{-1}$ and $\alpha'$ is dimensionless.

The outflowing gas travels at approximately the thermal speed, given by $V_{g} = (8 k T/(\pi \mu m_H))^{1/2}$, where $k = 1.38\times10^{-23}$ J K$^{-1}$ is  Boltzmann's constant, $\mu$ is the molecular weight and $m_H = 1.67\times10^{-27}$ kg is the mass of hydrogen.  Ices  like CO ($\mu$ =  28) are so volatile that their sublimation depresses the surface temperature to  $T \sim$ 20 to 25 K, approximately independent of heliocentric distance, at which temperature the thermal speed  is $V_{g} \sim$ 130 m s$^{-1}$.    The best-measured value of the momentum transfer coefficient  is $k_R$ = 0.5, for 67P/Churyumov-Gerasimenko (Appendix to Jewitt et al.~2020).  Measurements of other comets show that $f_{dg}$ is spread over a wide range,  with most of the mass carried by large particles like those in the coma of K2.   Unfortunately, neither $k_R$ nor $f_{dg}$ has been measured in comet K2.  For the sake of definiteness, we adopt $k_R$ =  0.5, $f_{dg} = 1$, $\rho$ = 500 kg m$^{-3}$  and substitute $dM/dt$ from Equation (\ref{dmbdt}) into Equation (\ref{accn}).  Neglecting a small residual heliocentric dependence, we obtain

\begin{equation}
\alpha' \sim 2\times10^{-4}  \left(\frac{\textrm{1 km}}{r_n}\right)^3 
\label{alpha'}
\end{equation}

\noindent with $r_n$ expressed in kilometers.

\begin{figure}
\epsscale{0.8}
\plotone{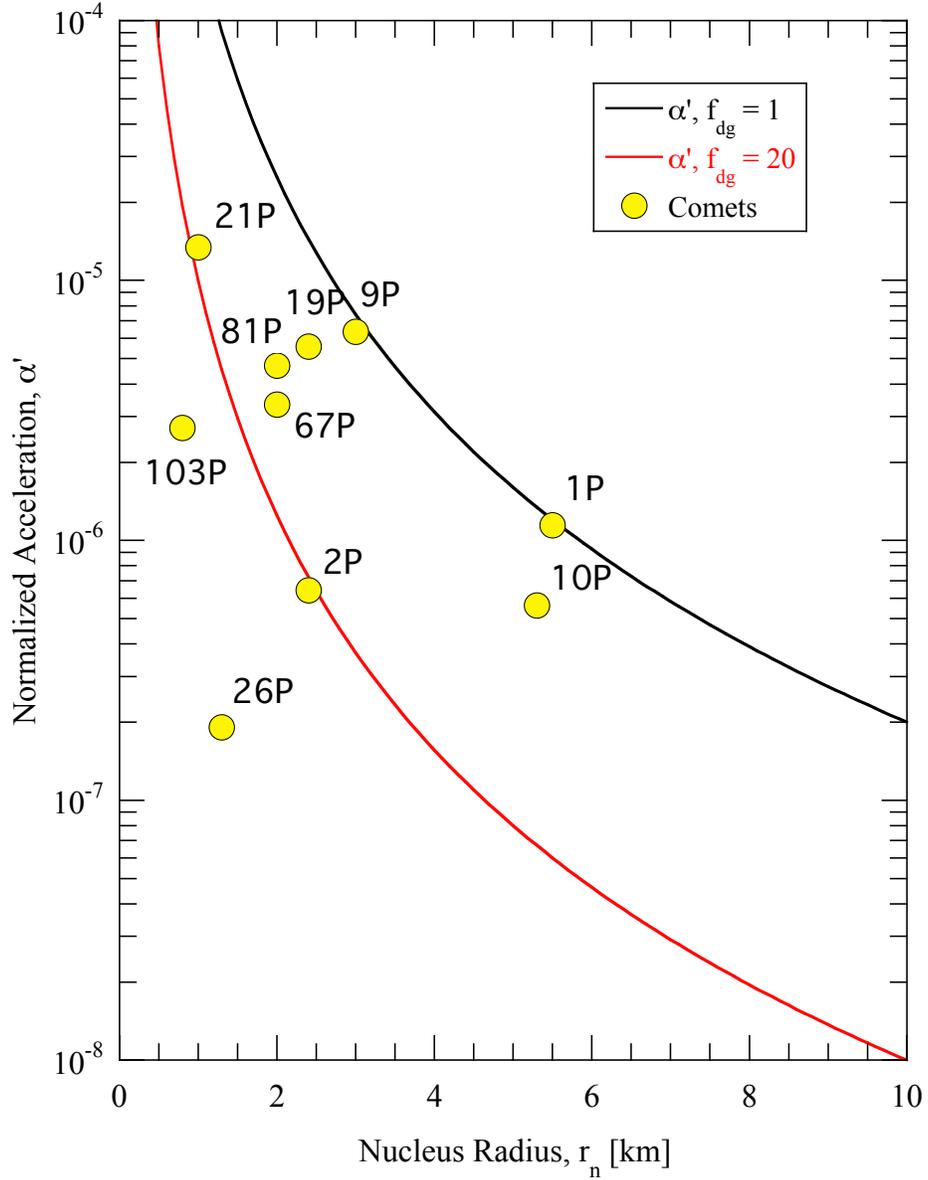}
\caption{Normalized non-gravitational acceleration of C/2017 K2  is plotted as a function of nucleus radius, $r_n$, from Equation (\ref{alpha'}).  Black and red curves show dust to gas ratios $f_{dg}$ = 1 and 20, respectively.  Yellow-filled circle symbols show, for comparison, peak (perihelion) normalized accelerations of  well-studied periodic comets, as labelled.  
\label{alpha'_vs_rn}}
\end{figure}

The solid black  and red lines in Figure (\ref{alpha'_vs_rn})  show $\alpha'$ as a function of $r_n$, for assumed values of the dust to gas ratio $f_{dg}$ = 1 and 20, respectively.  We have plotted $\alpha'$ in the radius range 0 $< r_n \le$ 10 km, bearing in mind the empirical upper limit to the radius of K2 ($r_n \le$ 9 km) set by examination of the surface brightness profile (Jewitt et al.~2019).  Also shown are individual peak values of $\alpha'$ for well-characterized cometary nuclei, using nucleus radii and solutions for the non-gravitational parameters $A_1$, $A_2$ and $A_3$ taken from NASA's JPL Horizons site\footnote{$\url{https://ssd.jpl.nasa.gov/horizons.cgi}$}.  
The magnitude of $\alpha'$ varies around the orbit, but reaches a maximum value near perihelion, where outgassing is strongest.  Therefore, we show $\alpha'$ computed with $r_H$ equal to the perihelion distance of each comet.  Specifically, we used

\begin{equation}
\alpha' = \frac{g(q) q^2}{G M_{\odot}} (A_1^2 + A_2^2 + A_3^2)^{1/2}
\label{alphadash}
\end{equation}

\noindent where  $g(q)$ is the function defined by Marsden et al.~(1973)  to represent the sublimation rate of water ice, evaluated at the perihelion distance, $q$.  We note that, in Equation (\ref{alphadash}) the total acceleration  is dominated by the radial component, $A_1$, because the bulk of the mass loss is directed sunward from the heated dayside of the nucleus.  The  resulting points plotted in Figure (\ref{alpha'_vs_rn}) should be regarded as upper limits to $\alpha'$ in each case because, unlike supervolatile CO, the sublimation rate of less volatile water ice falls faster with heliocentric distance than $r_H^{-2}$.  

Figure (\ref{alpha'_vs_rn}) shows that, depending on the nucleus radius, the dynamical effect of outgassing in K2, even when beyond Saturn, should be comparable to that in  comets of the inner solar system, for which $10^{-7} \lesssim \alpha' \lesssim 10^{-5}$.    However,  observations of short-period comets are possible over multiple orbits, making visible the cumulative effects of even very small $\alpha'$. The detection of non-gravitational acceleration in K2 (and other long-period comets for which only single-orbit observations are possible) will be more difficult.  For example, consider a nucleus with $f_{dg}$ = 1 and $r_n$ = 9 km, the maximum  possible radius estimated for K2 (Jewitt et al.~2019). Figure (\ref{alpha'_vs_rn}) indicates $\alpha' \sim 2\times10^{-7}$.  The resulting displacement of the nucleus caused by non-gravitational acceleration acting continuously over the three years from 2017 to 2020 would be an immeasurably small $\sim$ 60 km.  For a smaller nucleus, $r_n$ = 3 km, $\alpha'$ increases to $\sim 10^{-5}$ and the resulting displacement would increase to $\sim$ 2,000 km. At 10 AU, this would subtend $\lesssim$0.3\arcsec~in the plane of the sky, comparable to or slightly smaller than the best astrometry reported for comets, and therefore still very difficult to detect.  Moreover, in single-orbit observations, a non-zero $\alpha'$ can be easily misinterpreted as caused by an orbital eccentricity slightly different from the outgassing-free value.  For these reasons, non-gravitational accelerations of long-period comets remain undetected in the middle and outer solar system, at least in  comets with perihelia $q \gtrsim$ 3 AU (Krolikowska 2020).  

Two observational improvements are likely to change this circumstance.  First,  astrometric accuracy is being steadily improved by new all-sky astrometric surveys (especially from GAIA but also Pan STARRS and, soon, LSST).  Second, deeper surveys will lead to the discovery of even more distant long-period comets, bringing more attention to the dynamical effects of outgassing at Kuiper belt distances. Ultimately, the accuracy with which the properties of the Oort cloud can be deduced will depend on orbit determinations that incorporate outgassing forces over the full range of heliocentric distances.

Additional observations are planned to examine the development of K2 upon approach to the Sun.  Extrapolation of Equation (\ref{dmbdt}) to smaller heliocentric distances would suggest that K2 could become very bright.  We suspect, however, that such an extrapolation will fail, because the style of the outgassing will change as the distance shrinks.  For example,  an increase in the activity level and brightness might occur if buried amorphous ice is crystallized by the rising heat of the Sun.  On the other hand, icy grains in the coma, responsible for essentially all of the scattering cross-section of the comet, might soon begin to sublimate or thermally disaggregate, causing the brightness to fade.   Water ice, until now dormant in the nucleus and perhaps in the grains of the coma, will begin to sublimate at about 5 AU, changing the style of the activity and perhaps overwhelming the mass loss driven by supervolatile sublimation.  For all these reasons, we are reluctant to offer a prediction concerning the perihelion brightness of K2, while remaining excited to follow the development of this object across an unprecedented range of heliocentric distances.

\clearpage

\section{SUMMARY}
We present  Hubble Space Telescope measurements of inbound, long-period comet C/2017 K2 over the range of heliocentric distances from $r_H$ = 15.9 AU to 8.9 AU.  We find that

\begin{enumerate}

\item The particle properties established in previous observations (a coma of large, slowly-moving grains distributed in a nearly spherical coma) are unchanged across this heliocentric distance range.

\item  The dust scattering cross-section, measured within a fixed, nucleus-centered volume,  varies with heliocentric distance as $C_e \propto r_H^{-1.14\pm0.05}$.  This weak distance dependence reflects an inverse relation between dust ejection speed and distance and also implies that a significant fraction of the dust cross-section is carried by ultra-slow, nearly co-moving particles released at much larger distances.  

\item The release of ultra-slow particles began at Kuiper belt distances ($r_H \sim$ 35 AU), presumably driven by the sublimation of  carbon monoxide (or other supervolatile ice).


\item The normalized non-gravitational acceleration, $\alpha'$, even when  in the giant planet region of the solar system,  may rival $\alpha'$ measured in comets in the terrestrial planet region.  Distant outgassing may set the ultimate limit to the accuracy with which the orbits of long-period comets (and so the structure of the Oort cloud) can be deduced.

\end{enumerate}

\acknowledgments
We thank the anonymous referee for comments. Based on observations made under GO 14939, 15409, 15423 and 15973 with the NASA/ESA Hubble Space Telescope, obtained at the Space Telescope Science Institute,  operated by the Association of Universities for Research in Astronomy, Inc., under NASA contract NAS 5-26555.    Y.K. and J.A. acknowledge funding by the Volkswagen Foundation. J.A.'s contribution was made in the framework of a project funded by the European Union's Horizon 2020 research
and innovation programme under grant agreement No 757390 CAstRA.



{\it Facilities:}  \facility{HST}.

\clearpage


\clearpage

\begin{deluxetable}{lccrrrrrrr}
\tabletypesize{\scriptsize}
\tablecaption{Observing Geometry 
\label{geometry}}
\tablewidth{0pt}
\tablehead{\colhead{UT Date \& Time} & \colhead{Tel\tablenotemark{a}}  & \colhead{DOY\tablenotemark{a}} & \colhead{$r_H$\tablenotemark{b}} & \colhead{$\Delta$\tablenotemark{c}}  & \colhead{$\alpha$\tablenotemark{d}} & \colhead{$\theta_{- \odot}$\tablenotemark{e}} & \colhead{$\theta_{-V}$\tablenotemark{f}} & \colhead{$\delta_{\oplus}$\tablenotemark{g}}   }

\startdata

2017 Jun 28 20:09 - 20:52\tablenotemark{h} & HST &179 & 15.869 & 15.811 & 3.7 &  166.3  & 357.2  & +0.59 \\

2017 Nov 28 17:08 - 17:52 			&  HST &332 & 14.979 & 15.133 & 3.7&  17.1 & 358.3 & +1.38\\

2017 Dec 18 22:30 - 23:13 			&  HST &352 & 14.859 & 15.010 & 3.7 &  358.4 & 357.3  & +0.08 \\
2018 Mar 17 09:28 - 10:34 			&  HST &441 & 14.331 &14.328  & 2.0 &  275.6  & 353.3  & -3.97 \\
2018 Jun 15  15:40 - 16:19 			&  HST & 531 & 13.784 & 13.668 & 4.2 &  181.3 & 356.3 & -0.28 \\
2019 Oct 03 04:40 - 05:17            		& HST & 1007 & 10.736 & 10.871 & 5.3 & 70.9 & 2.1 & +5.15  \\
2019 Dec 17 01:33 - 02:10 			& HST & 1082 & 10.224 & 10.536 & 5.2 & 1.3 & 358.0 & +0.31  \\

2020 Mar 14 20:28 - 21:05 			& HST & 1169 & 9.603 & 9.634 & 5.9 & 278.8 & 352.5 & -5.94\\

2020 Jun 18 18:51 - 19:28 			& HST & 1265 & 8.913 & 8.626 & 6.4 & 178.1 & 357.3 & -0.08 \\

\enddata


\tablenotetext{a}{Day of Year, DOY = 1 on UT 2017 January 01}
\tablenotetext{b}{Heliocentric distance, in AU}
\tablenotetext{c}{Geocentric distance, in AU}
\tablenotetext{d}{Phase angle, in degrees}
\tablenotetext{e}{Position angle of projected anti-solar direction, in degrees}
\tablenotetext{f}{Position angle of negative heliocentric velocity vector, in degrees}
\tablenotetext{g}{Angle from orbital plane, in degrees}
\tablenotetext{h}{Observations from GO 14939, described in Jewitt et al.~(2017)}

\end{deluxetable}

\clearpage

\begin{deluxetable}{lccccccc}
\tabletypesize{\scriptsize}

\tablecaption{Fixed-Aperture Photometry\tablenotemark{a} 
\label{photometry}}
\tablewidth{0pt}

\tablehead{ \colhead{UT Date} & $\theta_5$\tablenotemark{b} & $\ell/10^3$ = 5 & 10  & 20 & 40 & 80 & 160}
\startdata
2017 Jun 28 	& 0.44			& 21.59/9.45/2.5 & 20.80/8.66/5.2 & 20.04/7.90/10.4	& 19.34/7.20/19.8 & 18.83/6.69/32 & 18.63/6.49/38  \\
2017 Nov 28  & 0.46                          & 21.31/9.38/2.6    & 20.54/8.61/5.4 & 19.78/7.85/10.9 & 19.09/7.16/20.4   &  18.64/6.71/31 & --- \\

2017 Dec 18   & 0.46                         & 21.31/9.42/2.6    & 20.52/8.63/5.3   & 19.74/7.85/10.9    & 19.03/7.14/20.8  & 18.53/6.64/33 & 18.27/6.38/42 \\

2018 Mar 17   & 0.48				& 21.03/9.31/2.8 & 20.26/8.54/5.8& 19.49/7.77/11.7 &  18.80/7.08/22.1  & 18.32/6.60/34 & 18.13/6.41/41\\
2018 Jun 15   & 0.50				& 20.81/9.27/2.9 & 20.05/8.51/5.9 & 19.29/7.75/11.9 & 18.57/7.03/23.1   & 18.02/6.48/38  & 17.78/6.24/48 \\
2019 Oct 03 	& 0.63		& 19.61/9.06/3.6 & 18.86/8.31/7.1 & 18.12/7.57/14.1 & 17.39/6.84/27.6 & 16.78/6.23/48 & 16.55/6.00/60\\

2019 Dec 17 & 0.65			&    19.38/9.01/3.7  & 18.63/8.26/7.4  & 17.90/7.53/14.6  & 17.18/6.81/28.3  & 16.57/6.20/50   & 16.31/5.94/63 \\

2020 Mar 14 & 0.72			&  19.02/8.95/3.9 & 18.27/8.20/7.8  & 17.53/7.46/15.5  & 16.83/6.76/29.6  & 16.23/6.16/51  & 15.95/5.88/66  \\

2020 Jun 18 	& 0.80		&  18.56/8.88/4.2  & 17.80/8.11/8.5  & 17.06/7.38/16.8  & 16.35/6.67/32.4 & 15.72/6.04/58  & 15.43/5.75/76   \\

\enddata

\tablenotetext{a}{For each date and aperture radius, $\ell$ (measured in units of 10$^3$ km at the comet), the Table lists the apparent magnitude, V, the absolute magnitude, H, and the scattering crossection, $C_e$ (in units of 10$^3$ km$^2$), in the order V/H/$C_e$. $C_e$ is computed from H using Equation (\ref{area}).}
\tablenotetext{b}{$\theta_5$ is the angle subtended by the radius of the smallest (5000 km) aperture, in arcsecond.}

\end{deluxetable}

\clearpage

\begin{deluxetable}{llcllcc}
\tabletypesize{\scriptsize}

\tablecaption{Size Distribution Indices\tablenotemark{a} 
\label{indices}}
\tablewidth{0pt}

\tablehead{ \colhead{Comet} & Method\tablenotemark{b} & Radii ($\mu$m) & Index, $q$ & Reference }
\startdata

1P/Halley								& In-Situ & $>$20     & 3.5$\pm$0.2 		  &  Fulle et al.~(1995) \\

2P/Encke								& Optical &  $>$ 1 & 3.2 to 3.6 			    & Sarugaku et al.~(2015) \\

22P/Kopff 								& Optical & $>$ 1  & 3.1				   & Moreno et al.~(2012) \\

26P/Grigg-Skjellerup						& Optical & $>$ 60 & 3.3			            & Fulle et al.~(1993) \\

67P/Churyumov-Gerasimenko (coma)		& In-Situ & $>$ 0.01 & 3.7$_{-0.1}^{+0.7}$  & Marschall et al.~(2020)  \\
67P/Churyumov-Gerasimenko (trail)			& Optical & $>100$   & 4.1			  & Agarwal et al.~(2010) \\

81P/Wild								& Optical & $>$ 1 & 3.45$\pm$0.1 		  & Pozuelos et al.~(2014) \\

103P/Hartley 							& In-Situ & $>$ 10$^4$ & 4.7 to 6.6 		  & Kelley et al.~(2013) \\
103P/Hartley 							& Optical & $>$ 1 & 3.35$\pm$0.1 		  & Pozuelos et al.~(2014) \\

209P/LINEAR							& Optical & $>$ 1  & 3.25$\pm$0.1              & Ishiguro et al.~(2015) \\

\enddata

\tablenotetext{a}{Reported differential power-law size distribution index, $q$.}
\tablenotetext{b}{In-Situ: measured by a spacecraft in the coma.  Optical: remote determination by fitting tail isophotes}

\end{deluxetable}

\clearpage

\clearpage

\clearpage

\clearpage

\clearpage

\clearpage

%

\clearpage

\clearpage

\clearpage

\clearpage

\clearpage

\end{document}